\documentclass[aps,twocolumn,pra,superscriptaddress]{revtex4-2}

\usepackage[english]{babel}
\usepackage{graphicx}
\usepackage{dcolumn}
\usepackage{bm}
\usepackage{physics}
\usepackage{amsmath,amsfonts,amssymb}
\usepackage{siunitx}
\usepackage{xcolor}

\begin{document}

\preprint{APS/123-QED}

\title{Acceleration-driven dynamics of \textcolor{black}{Josephson vortices} in coplanar superfluid rings
}


\author{Yurii Borysenko}
\affiliation{Department of Physics, Taras Shevchenko National University of Kyiv, 64/13, Volodymyrska Street, Kyiv 01601, Ukraine}

\author{Nataliia Bazhan}
\affiliation{Vienna Center for Quantum Science and Technology, Atominstitut, TU Wien,
Stadionallee 2, 1020 Vienna, Austria}

\author{Olena Prykhodko}
\affiliation{Department of Physics, Taras Shevchenko National University of Kyiv, 64/13, Volodymyrska Street, Kyiv 01601, Ukraine}

\author{Dominik Pfeiffer}
\affiliation{Technische Universit\"at Darmstadt, Institut f\"ur Angewandte Physik, Schlossgartenstraße 7, 64289 Darmstadt, Germany}

\author{Ludwig Lind}
\affiliation{Technische Universit\"at Darmstadt, Institut f\"ur Angewandte Physik, Schlossgartenstraße 7, 64289 Darmstadt, Germany}

\author{Gerhard Birkl}
\affiliation{Technische Universit\"at Darmstadt, Institut f\"ur Angewandte Physik, Schlossgartenstraße 7, 64289 Darmstadt, Germany}
\affiliation{Helmholtz Forschungsakademie Hessen für FAIR (HFHF), Campus Darmstadt, Schlossgartenstraße 2, 64289 Darmstadt, Germany}

\author{Alexander Yakimenko}
\affiliation{Department of Physics, Taras Shevchenko National University of Kyiv, 64/13, Volodymyrska Street, Kyiv 01601, Ukraine}
\affiliation{Dipartimento di Fisica e Astronomia Galileo Galilei, Universit\'a di Padova, and INFN, Sezione di Padova,
Via Marzolo 8, 35131 Padova, Italy}


\date{\today}

\begin{abstract}
Precise control of topologically protected excitations, such as quantum vortices in atomtronic circuits, opens new possibilities for future quantum technologies.
We theoretically investigate the dynamics of Josephson vortices (rotational fluxons) induced by coupled persistent currents in a system of coplanar double-ring atomic Bose-Einstein condensates. 
We study the Josephson effect in an atomic Josephson junction formed by coaxial ring-shaped condensates. Tunneling superflows, initiated by an imbalance in atomic populations between the rings, are significantly influenced by the persistent currents in the inner and outer rings. 
This results in pronounced Josephson oscillations in the population imbalance for both co-rotating and non-rotating states. 
\textcolor{black}{If a linear acceleration is applied to the system, our analysis reveals peculiar azimuthal tunneling patterns and dynamics of Josephson vortices  which leads to non-zero net tunneling \textcolor{black}{current} and shows sensitivity to the acceleration magnitude.
When multiple Josephson vortices are present, asymmetric vortex displacements that correlate with both the magnitude and direction of acceleration can be measured, offering potential for quantum sensing applications.
}
\end{abstract}



\maketitle



\section{Introduction}
The Josephson effect (JE), first predicted in superconductors \cite{Josephson1962, Barone1982}, has since been observed in a wide range of different types of Josephson junctions. Ring-shaped geometries are of particular interest, as they support stable supercurrents, commonly known as persistent currents. Cylindrical superconducting long Josephson junctions and fluxons have been the focus of extensive research over the past several decades \cite{Tilley1966, Burt1981, Sherrill1979, Davidson1986, Ustinov1992, Hermon1994, Ustinov1996, Ustinov1996, Ustinov1999, Watanabe1996, Trias2000, Ustinov2002, Ustinov2004, Fistul2003, Watanabe1996, Trias2000}.

The observation of alternating current (AC) and direct current (DC) Josephson effects in Bose-Einstein condensates (BECs) by \cite{Levy2007}, along with subsequent studies on bosonic Josephson junctions by \cite{Pigneur2018}, has further stimulated theoretical and experimental investigations of Josephson effects in quantum gases.

Persistent currents in toroidal atomic BECs have been the focus of extensive theoretical and experimental investigations as a signature of superfluidity at \textcolor{black}{macroscopic} scale \cite{PhysRevLett.99.260401,PhysRevLett.111.205301,2020NatCo..11.3338R,PhysRevA.86.013629,PhysRevA.80.021601,PhysRevA.81.033613,PhysRevLett.106.130401,2001JPhB...34L.113B,PhysRevA.74.061601,PhysRevLett.110.025301,PhysRevA.88.051602,
tononi2024quantum, Polo2024}. 
The toroidal trap geometry, characterized by a substantial central hole surrounding the axis of the condensate, inherently bounds the core of the vortex states within the effective potential trap. 
This confinement enhances the stability even for multicharged vortices. 
The robustness of persistent currents in a single ring  
\textcolor{black}{naturally leads to the exploration of quantized angular momentum}
in two parallel-coupled superfluid rings, particularly concerning the JE in such a dual-ring configuration 
\cite{Lesanovsky2007,
Brand2009,
Brand2010,
Brand2013,
2012JPhB...45c5004Q,
2018ScPP....4...18B,PhysRevA.98.053603,PhysRevLett.111.105302,PhysRevA.93.033618,sym11101312,OLIINYK2020105113,Oliinyk2019,Bazhan2022,hernandez2024connecting,pezze2024stabilizing}.





\textcolor{black}{A key phenomenon in such systems is the formation of Josephson vortices (JVs). These vortices, also known as rotational fluxons, result from the phase difference across a junction between two BECs. As topologically protected structures, they are very robust and valuable for quantum sensing applications \cite{Kuklov2005,Kuklov2006}. }  In atomic BEC systems, JVs are particularly important for exploring new applications in atomtronics, where their unique properties can be used for building quantum devices \cite{amico:2017:atomtronics,amico:2021:roadmap}.

\textcolor{black}{Very recently, a side-by-side configuration of two rings connected by a tunable weak link was proposed as a physical platform for creating acceleration \cite{Dimer_accelerated} and rotation \cite{edwards2024target} sensors, based on a threshold-driven vortex transfer approach previously introduced in \cite{Bland22}. In this setup, the barrier amplitude directly modulates vortex transitions, enabling discrete, measurable shifts that can be finely controlled, or even halted, by tuning the barrier strength.}



\begin{figure}[htb]
    \centering \includegraphics[width=\columnwidth]{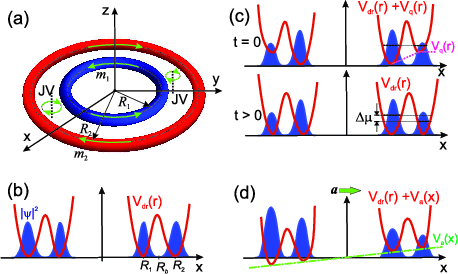}
\caption{ (a) Schematic of the coplanar double-ring BEC with counter-propagating superflows (green arrows) and \textcolor{black}{JVs} (black dotted lines). (b) Potential profile $V_{dr}(r)$ (red) along the $x$-axis, forming the double-ring trap and condensate density $|\psi|^2$ (blue). (c) Initial state at $t=0$ showing \textcolor{black}{creation of} chemical potential difference $\Delta \mu$ via tilt potential $V_q(r)$ (magenta), quenched to the symmetric state for $t>0$. (d) Uniform linear acceleration along the $x$-axis (green arrow), with effective potential $V_a(x)$ (green dash-dotted line) inducing a density gradient. }    \label{sketch}
\end{figure}

In the present work, we investigate the dynamics of JVs in a dual co-planar ring configuration of atomic BECs (see Fig.\ref{sketch}).
First, we study the AC JE in coaxially aligned toroidal condensates separated by a potential barrier. 
We demonstrate that tunneling superflows, initiated by an imbalance in atomic population between the rings, are strongly influenced by the persistent currents in both the inner and outer rings. This leads to pronounced Josephson oscillations in the population imbalance for rings with identical angular momentum states. In contrast, rings with different angular momentum states exhibit zero net current across the junction. We analyze the azimuthal pattern of the tunneling flow and JVs in the circular junction between rings. Furthermore, we investigate the impact of linear acceleration on vortex dynamics, showing that it induces an asymmetric displacement of the JVs. This asymmetry in JV position can be quantified and used to determine both the magnitude and direction of the acceleration.

The article is organized as follows. In Section II we analyze Josephson oscillations and the dynamics of JVs in a double-ring system. We highlight several different dynamical regimes of the system depending on the topological charges of the rings and the magnitude of acceleration. Section III examines the relaxation dynamics of JVs under linear acceleration. It also demonstrates how the stabilized arrangement of Josephson vortices between the rings can be used to measure the direction and magnitude of the external acceleration. Section IV provides summary and conclusions.


\section{Josephson oscillations in double-ring system}

The dynamical properties of a BEC within the mean-field theory at the zero-temperature limit is governed by the Gross-Pitaevskii equation (GPE):
\begin{equation}
i\hbar\frac{\partial\Psi(\mathbf{r},t)}{\partial t}= \left[-\frac{\hbar^{2}}{2M}\nabla^{2}+V_\textrm{ext}(\mathbf{r})+{g}|\Psi(\mathbf{r},t)|^{2}\right]\Psi(\mathbf{r},t). 
\end{equation}
where  $g=4\pi a_{s}\hbar^{2}/M$, $M=\SI{1.44e-25}{\kilo\gram}$, and  $a_{s} = \SI{5.3e-9}{\meter}$ are the interaction strength, mass, and the $s$-wave scattering length of $^{87}$Rb atoms. The wave function is normalized to the total number of atoms in the system: 
\begin{equation}
\int \left|\Psi(\mathbf{r},t)\right|^2 d \mathbf{r} = N.
\end{equation}
The external trap potential $V_\textrm{ext}$ form the double-ring geometry of the system (see Fig.\ref{sketch}):
\begin{equation}
    V_\textrm{ext}(\mathbf{r}) = \frac{1}{2}M\omega_z^2 z^2 + V_{dr}(r),
\end{equation}
where  $\omega_z=2\pi\times\SI{245}{\hertz}$ is the trapping frequency in a light sheet confining the atoms along the $z$ direction.
The radial trapping potential is 
\begin{equation}
    V_{dr}(r) = V_1(r)\Theta(R_1-r)+V_2(r)\Theta(r-R_2)+V_b(r),
\end{equation}
where $r=\sqrt{x^2+y^2}$, $\Theta(r)$ is the Heaviside theta-function, and
$V_j(r)$ for $j=1,2$ is given as
\begin{equation}
V_{j}(r)=\frac{1}{2}M\omega^{2}_{r}(r-R_{j})^2, 
\end{equation}
with equal trapping frequencies  $\omega_r = 2\pi\times\SI{110}{\hertz}$ and different ring radii $R_{1}=\SI{14}{\micro\meter}$, $R_{2}=\SI{24}{\micro\meter}$. 
The potential 
$V_b(r)$ describes the barrier separating the rings as 
\begin{equation}
V_b(r)=U_b\exp\left[-\frac{(r-R_b)^2}{2l_{b}^2}\right],    \end{equation}
centered at $R_b=(R_1+R_2)/2$. The barrier height $U_b$ and width $l_b$ are specified below.


\textcolor{black}{We analyze tunneling flows based on the full two-dimensional (2D) GPE for a setup of co-planar rings, assuming tight confinement along the $z$-axis, so that bending and tilting of JV are suppresssed, allowing for a 2D approximation:}
\begin{equation}
    \Psi(\mathbf{r},t) = \psi(x,y,t)\zeta(z,t),
\end{equation}
with 
\begin{equation}
    \zeta(z,t) = \left(\frac{1}{\sqrt{\pi}l_{z}}\right)^{1/2} \exp\left(-\frac{z^{2}}{2l_{z}^2}-i\frac{\omega_z t}{2}\right), 
\end{equation}
where 
$l_{j} = \sqrt{\hbar/(M\omega_{j})}$, $j=z,r$.
Applying the following transformations  $t \to \omega_r t$, $(x,y,z)\to (x, y, z)/l_r$, $V\to V/(\hbar\omega_r)$, $\mu\to \mu/(\hbar\omega_r)$, $\psi \to l_r^{3/2}  \psi$, the dimensionless 2D GPE can be written as:
\begin{equation}\label{GPE2Ddimless}
     i\frac{\partial \psi }{\partial t}= \left(-\frac{1}{2}
 \nabla ^{2}+V_\textrm{ext}+g_{\textrm{2D}}|\psi |^{2} -\mu \right) \psi, 
\end{equation}
where $g_{\textrm{2D}} = g/\sqrt{2\pi}l_z$ is the dimensionless coupling and $\mu$ is the chemical potential of the steady state $\tilde{\psi}$, which satisfies the stationary GPE:  $\hat{\mathcal{H}}\tilde{\psi}=\mu\tilde{\psi}$
with
\begin{equation}
\hat{\mathcal{H}} = -\frac{1}{2}
 \nabla ^{2}+V_\textrm{ext}+g_{\textrm{2D}}|\tilde{\psi}|^{2}.
    \label{eq:HatH}
\end{equation}

The wave function $\Tilde{\psi}(x,y)$ of the stationary state of the condensate in the double-ring potential $V_{dr}(r)$ is obtained by the imaginary time propagation method. 
The total number of atoms $N=N_1^{(0)}+N_2^{(0)}$ is distributed between inner and outer rings:
\begin{equation}
N_j^{(0)}=\int\int_{S_j}|\Tilde{\psi}|^2dxdy
    \label{eq:Nequlibrium}
\end{equation}
with integration boundaries for the inner ring, $S_1$: ${0\le r<R_b}$ and for the outer ring, $S_2$: ${r \ge R_b}$. 
The radial double-well potential $V_{dr}(r)$ traps an azimuthally-symmetric double-ring condensate and splits it into two parts.




We define the population imbalance between these parts as the deviation in the number of particles from their equilibrium values, $N_j^{(0)}$, as follows:
\begin{equation}
 \Delta N(t) = \left[N_2(t)-N_2^{(0)}\right] - \left[N_1(t)-N_1^{(0)}\right].  
    \label{eq:Delta_N}
\end{equation}

The initial population imbalance and chemical potential difference $\Delta\mu=\mu_1-\mu_2$ between the rings is generated by applying an additional tilt potential, using methods similar to those employed in experiments with double-well systems \cite{Levy2007}. The total external potential comprises both the trapping potential and the tilt potential, as illustrated in Fig. \ref{sketch} (c). Here
$V_q(r)$ is a tilt potential
\begin{equation}
V_q(r) =
\left\{
\begin{array}{ll}
    b_1, & 0 \leq r < R_1, \\
    (b_{2}-b_{1})\frac{r-R_1}{R_2-R_1} + b_1, & R_1 \leq r < R_2, \\
    b_2, & R_2 \leq r,
\end{array}
\right.
\end{equation}
where $b_{1,2}$ are the potentials of the two rings biases, introduced to create a chemical potential difference $\Delta \mu$.
In our simulations we prepare the initial state of the system as a stationary state with $b_1=0$, $b_{2} \neq 0$, shifting up the outer ring, hence controlling $\Delta\mu$ by changing $b_2$. 
Then, to initiate the dynamics, we rapidly switch off the tilt potential observing the further evolution of the condensate in a double-ring system using conservative GPE. 

\begin{figure}
    \centering
\includegraphics[width=1.0\linewidth]{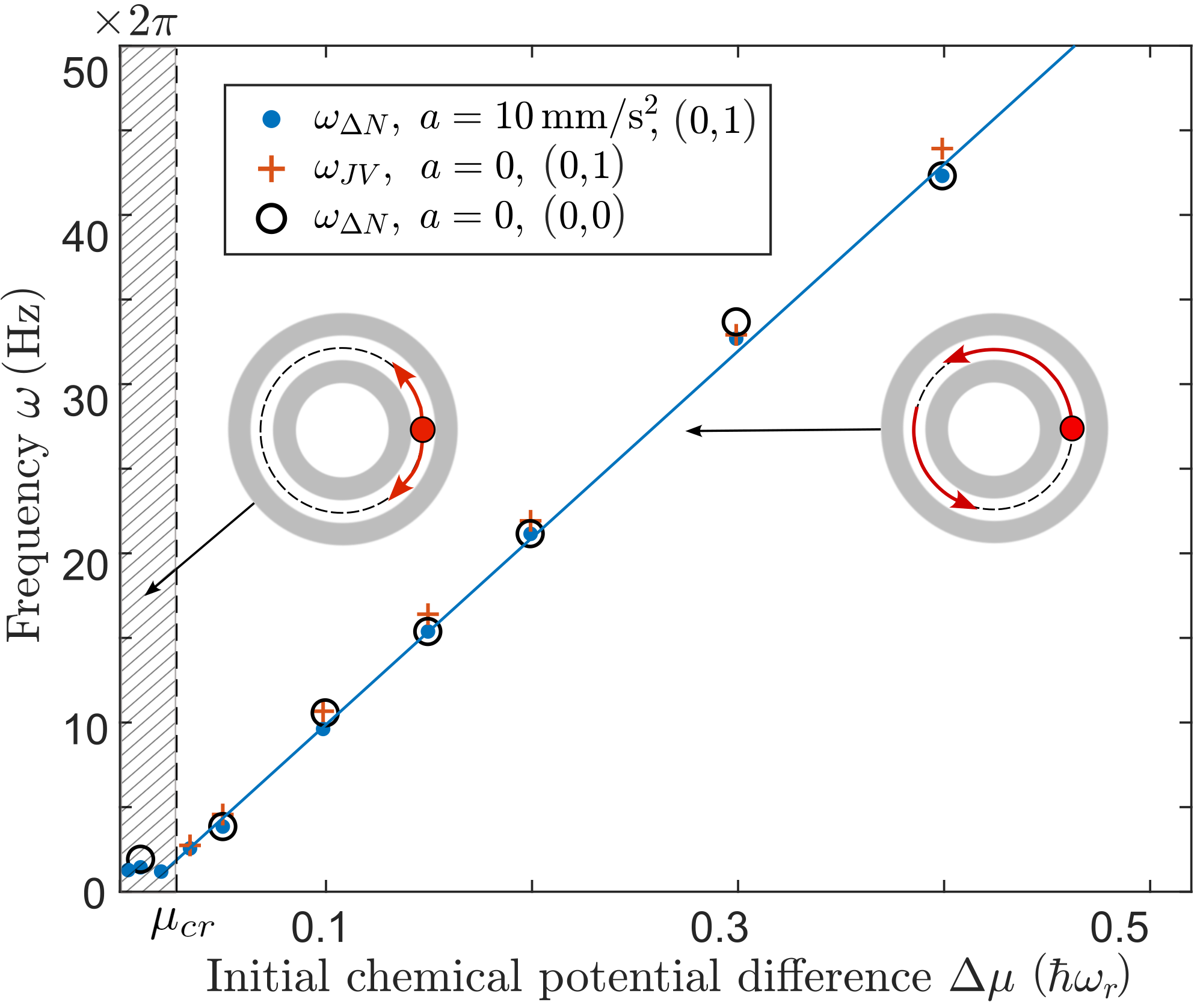}
    \caption{Oscillation frequency $\omega$ of the population imbalance vs. chemical potential difference $\Delta\mu$ for $(m_1 = 0, \, m_2 = 1)$ with $a = \SI{10}{\milli\meter/\second^2}$ (blue dots) and $(m_1 = m_2 = 0)$ without acceleration (black circles), fitted linearly. JV rotation frequencies for $m_1=0$, $m_2=1$ state are marked by red crosses.
    $U_b = 7\,\hbar\omega_r$, $l_b = \SI{1}{\micro\meter}$, $N = 5 \times 10^4$.}
    \label{JE_Freq_vs_dmu}
\end{figure}

First, we investigate the AC JE in a double-ring system, characterized by a constant chemical potential difference that drives oscillatory tunneling of atoms between the rings. 
A distinctive feature of the AC JE is that the frequency of the population imbalance is directly proportional to the applied chemical potential difference, while the phase difference increases linearly over time: $\Phi(t) = \Phi_0 + \omega_{\Delta N} t$. 
In our simulations the population imbalance oscillates with a frequency $\omega_{\Delta N} = \Delta \mu / \hbar$ for high enough initial chemical potential difference \textcolor{black}{(see Fig. \ref{JE_Freq_vs_dmu})}.  
These properties of the AC JE are observed in both non-rotating states and states with persistent currents, provided the rings have the same angular momentum state ($m_1 = m_2$).

\begin{figure}
    \centering
\includegraphics[width=1.0\linewidth]
{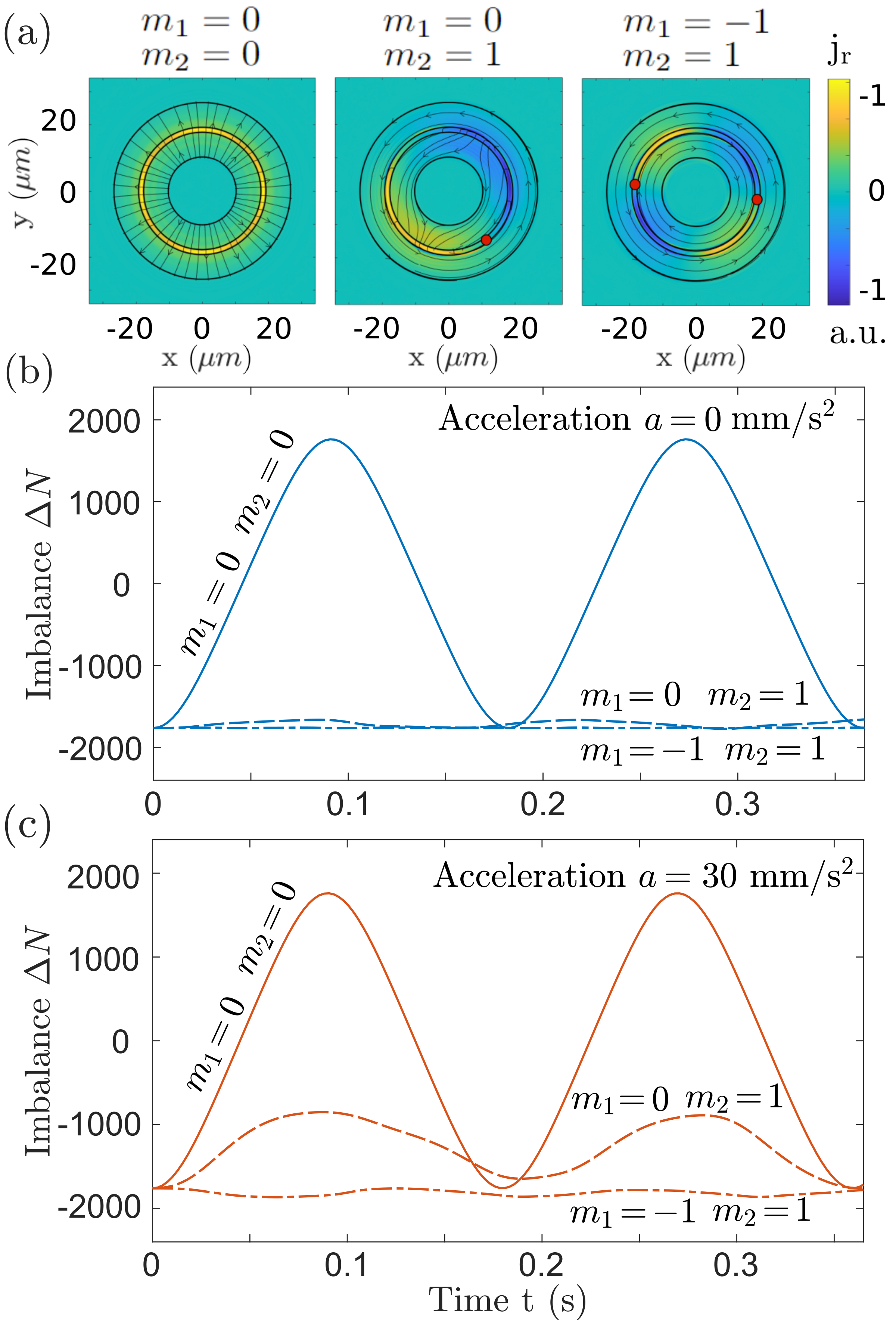}
\caption{(a) Flow density snapshots for different angular momentum states: $m_1 = m_2 = 0$ shows purely radial tunneling; $m_1 = 0$, $m_2 = 1$ forms one JV, highlighted by a red circle; $m_1 = -1$, $m_2 = 1$ results in two JVs. (b) Particle number imbalance for $m_1 = 0$, $m_2 = 0$ (solid blue, \textcolor{black}{pronounced tunneling}), $m_1 = 0$, $m_2 = 1$ and $m_1 = -1$, $m_2 = 1$ (dashed and dash-dotted blue, suppressed tunneling). (c) Linear acceleration induces imbalance oscillations for $m_1 = 0$, $m_2 = 1$ (dashed red) but not for counter-propagating flows (dash-dotted red). Tunneling for $m_1 = 0$, $m_2 = 0$ \textcolor{black}{is preserved} (solid red).
$U_b=49\,\hbar\omega_r=4.25\,\mu_0$, $l_b=\SI{0.33}{\micro\meter}$, $N = 5 \times 10^5$.}
\label{JE_both_n_from_t}
\end{figure}

We conducted an extensive series of numerical simulations of Josephson oscillations, varying the number of particles in the rings and the initial chemical potential difference. 
Figure \ref{JE_both_n_from_t} showcases pronounced oscillations and the radial flow structure for $N = 5 \times 10^5$. 
Figure \ref{JE_both_n_from_t} (b) specifically illustrates the $\Delta N(t)$ oscillations observed in the time-dependent GPE simulations for non-rotating rings ($m_1 = m_2 = 0$).

Remarkably, persistent currents with $m_1 \ne m_2$ result in a nearly constant $\Delta N(t)$ with no visible oscillations, as depicted in Fig. \ref{JE_both_n_from_t} (b). 

Numerical solutions of the GPE were used to calculate the superfluid-flow density as
\begin{equation}
    \mathbf{j} = \frac{i\hbar}{2M} \left[ \Psi \nabla \Psi^* - \Psi^* \nabla \Psi \right].\label{eq:current}
\end{equation}
Figure \ref{JE_both_n_from_t} (a) illustrates the radial flow distribution $j_r$ in the $(x, y)$ plane for different angular momentum states. 
Note that the suppression of the total tunneling flow is accompanied by the formation of $|m_1 - m_2|$ JVs, indicated by red circles in Fig. \ref{JE_both_n_from_t} (a). 
This phenomenon was previously observed in a system of vertically stacked ring-shaped condensates in Ref. \cite{Oliinyk2019} and attributed to the azimuthal symmetry of the tunneling flow. 
These radial flow properties are illustrated in Fig. \ref{JE_both_n_from_t} (a) for the planar double-ring system considered here.

Next, let us consider the effect of acceleration which breaks the azimuthal symmetry and causes a redistribution of density and radial flows. 
This additional density gradient can restore oscillations in $\Delta N(t)$, thereby activating Josephson oscillations driven by the applied acceleration. 
In our simulations, we introduce acceleration [Fig.~\ref{sketch} (d)] by adding   an effective potential of the form:
\begin{equation}
    V_a = M(\mathbf{a} \cdot \mathbf{r}) = M a x,
    \label{eq:Va}
\end{equation}
which corresponds to a linear horizontal acceleration $\mathbf{a} = (a, 0, 0)$. The introduction of such an effective potential corresponds to a transformation to an accelerating frame.

We prepare the initial state with the applied effective potential (\ref{eq:Va}) and introduce the chemical potential difference using the tilting potential procedure described above. 
The results of numerical simulations for non-zero acceleration are shown in Figs.~\ref{JE_both_n_from_t} (c) and \ref{JE_Freq_vs_dmu}. 
It is important to highlight the azimuthal symmetry-breaking effect induced by constant linear acceleration. 

To gain deeper insight into the tunneling flow mechanism between the condensate rings, we adopt a simplified two-mode approximation model. In this approach, we substitute the following ansatz for the 2D wave function into Eq.~(\ref{eq:current}): 
\begin{equation} \Psi = \psi_1(r)e^{i\mu_1 t + i m_1 \varphi} + \psi_2(r)e^{i\mu_2 t + i m_2 \varphi}, \label{eq:2_mode_WF} 
\end{equation} 
where $\psi_{1,2}$ are the wave functions of the condensate in the rings, $\mu_{1,2}$ are their chemical potentials, and $m_{1,2}$ represent the vorticities of the rings. Through straightforward algebra, we obtain very simple relation for the radial flow density: 
\begin{equation} j_r(r,\varphi,t) = j_0 \sin(\Delta\mu t - \Delta m \varphi), \label{eq:jr} 
\end{equation}
where $\Delta m = m_1 - m_2$ and $j_0 = j_0(r)$ is a function of the radial coordinate only, assuming an azimuthally symmetric density distribution ($a=0$). The effect of acceleration-induced density variations can be taken into account by modifying the amplitude factor as $j_0(r)\left(1 - \delta_a \cos\varphi\right)$, where $\delta_a \ll 1$ is a dimensionless parameter that accounts for the small density bias introduced by linear acceleration.

It is evident that the total flow through the junction, $J(t) = \int j_r\, d^2\mathbf{r}$, vanishes if the number of JVs is not zero or one. 
Specifically, $J(t)=J_0 \sin(\Delta \mu t)$ for $\Delta m = 0$, corresponding to AC Josephson oscillations with the frequency defined by the chemical potential difference. 
Remarkably, for $\Delta m = \pm 1$, the total tunneling flow is also non-zero: $J=\frac{1}{2} \delta_a J_0 \sin(\Delta \mu t)$, although the amplitude of the flow is reduced to the small factor $\delta_a$. 
These simple estimates align well with the results of numerical simulations, as illustrated in Fig. \ref{JE_both_n_from_t}. 

The dependence of the JE oscillation frequency $\omega_{\Delta N}$ on the initial chemical potential difference $\Delta\mu$ is depicted in Fig.~\ref{JE_Freq_vs_dmu}. 
Here and further we use a barrier width of $l_b=1\,\mu$m and the total particle number of $N=5 \times 10^4$.
Figure~\ref{JE_Freq_vs_dmu} shows the frequency of population imbalance oscillations, $\omega_{\Delta N}$ for two cases: (i) open circles depict  $a = 0$, $\Delta m = 0$, and (ii) blue dots correspond to $a = 10\,\textrm{mm}/\textrm{s}^2$, $\Delta m = 1$. 
The red crosses indicate the angular frequency $\omega_{JV}$ of the JV circulation in the barrier. 
Note that all three data sets fit the same linear behavior $\omega_{\Delta N} \sim \Delta\mu$ inherent to the AC JE for $\Delta\mu>\mu_{cr}$.

Figure~\ref{JE_Freq_vs_dmu} also reveals the two relevant regimes of JV motion depending on the additional energy provided by the initial chemical potential difference $\Delta\mu$, separating at $\Delta\mu = \mu_{cr}$. 
The frequency of Josephson oscillations is a linear function of the chemical potential difference if the initial imbalance is above a certain critical value $\mu_{cr}$. 
This regime corresponds to the circular motion of the JV within the barrier. 
The frequency $\omega_{JV}$ of the JV's circular motion in the barrier matches the frequency of the population imbalance $\omega_{\Delta N}$. 
Below the critical value $\mu_{cr}$ of the chemical potential difference, the frequency remains constant, corresponding to periodic oscillations of the JV around the equilibrium position with constant frequency.

\begin{figure}
    \centering
\includegraphics[width=1.0\linewidth]{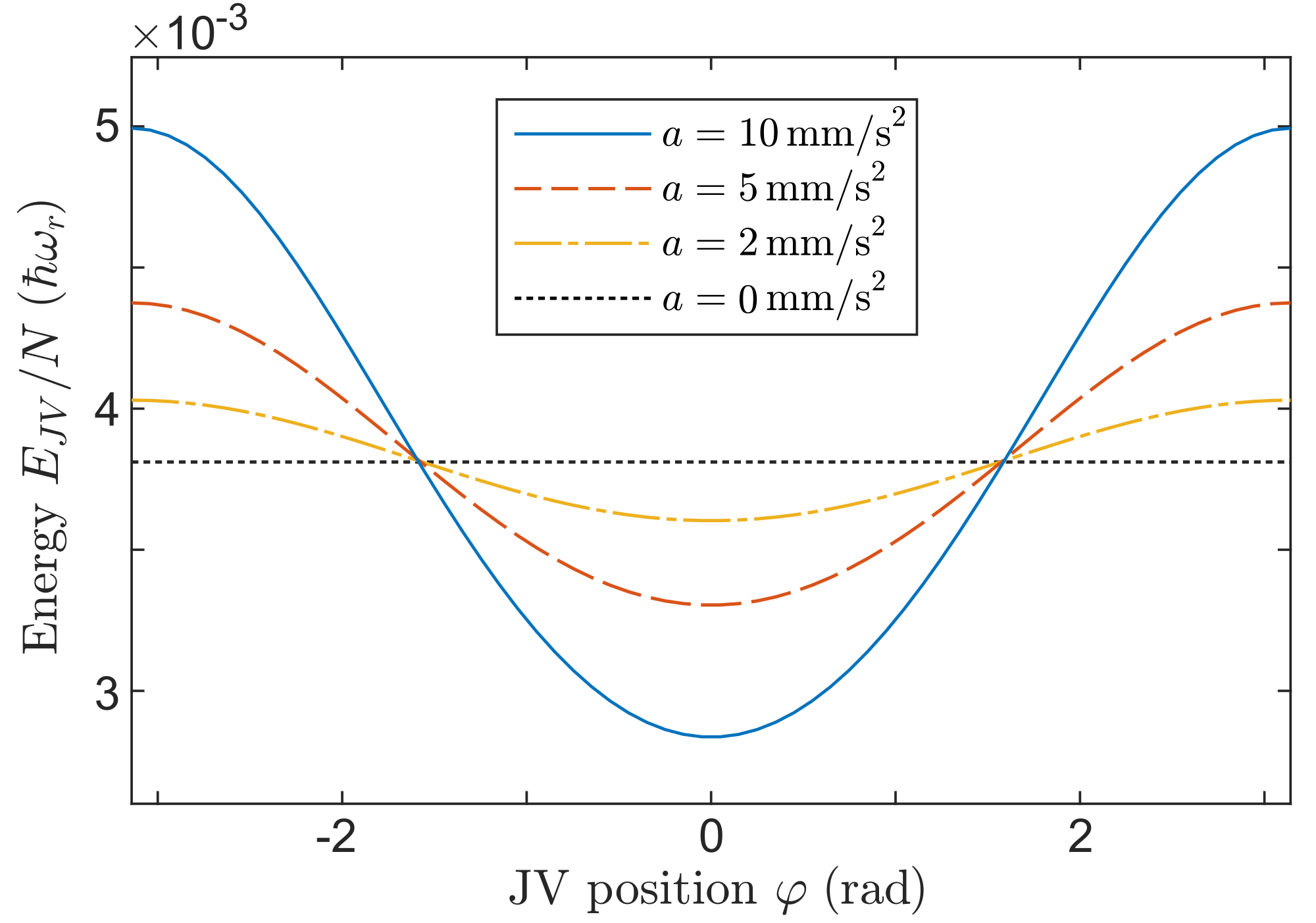}
    \caption{Nucleation energy per particle $E_{JV}/N$ for the single JV between rings as a function of JV's angular position $\varphi$. The acceleration-induced density gradient generates an effective potential well for the JV, with a minimum in the direction of the acceleration, \textcolor{black}{i.e., $\varphi_a = 0$}. 
    $U_b/\mu = 1.5$, $l_b = \SI{1.43}{\micro\meter}$, $N = 5 \times 10^4$.}
    \label{JV_Energy_analysis}
\end{figure}

We have used the energetic analysis for the JV position along the annulus (Fig.~\ref{JV_Energy_analysis}) and have found two distinct regimes of the JV dynamics: (i) circular motion in the barrier and (ii) oscillations in the vicinity of the position with minimum energy.
Let us show the connection between these characteristics and an external linear acceleration. 
To this end, we use the dependence of the BEC energy on the azimuthal position of the vortex in the ring.

Within the approximation of the constant local density of the unperturbed condensate (state $m_1=m_2=0$, $\psi_0$) in the vicinity of the vortex position, we can write the wavefunction of the state with imprinted vortex ($m_1=0,\:m_2=1$, $\psi_{\textrm{v}}$) through the wavefunction of the vortex in the homogeneous condensate and the unperturbed state $\psi_0$ as
\begin{equation}    \psi_{\textrm{v}}(\mathbf{r})=A\psi_0(\mathbf{r})\tanh{\bigg(\frac{\varrho}{\xi}\bigg)}e^{i\theta}
    \label{eq:JV_wavefunction}
\end{equation}
where $\varrho=|\mathbf{r}-\mathbf{r}_{JV}|$ is the distance from the vortex core located in $\mathbf{r}_{JV}$, $\theta=\arg{(\mathbf{r}-\mathbf{r}_{JV})}$ is the angle relative to the core, $A$ is the normalisation constant and $\xi$ is the healing length. 
Since the vortex is localised radially at the centre of the barrier, $r_{JV}=R_b$, thus its position and energy are uniquely determined by the angular position $\varphi_{JV}$. 

The total energy of the BEC in the state $\psi$ is given as follows: 
\begin{equation}
    E(\psi) = \int{\bigg( \frac{1}{2}\left|\nabla\psi\right|^2 + V_{\textrm{ext}}\left|\psi\right|^2 + \frac{g}{2}\left|\psi\right|^4 \bigg)\,d^2\mathbf{r}}. 
    \label{eq:JV_energy}
\end{equation}
The external potential $V_\textrm{ext}$ consists of both the trapping potential and the effective potential associated with the applied acceleration \eqref{eq:Va}.

Let us introduce a nucleation energy of the JV, defined as the difference
of the energy of the state with imprinted Josephson vortex line $\psi_{\textrm{v}}$ and the
ground-state energy 
\begin{equation}
E_{JV} = E(\psi_\textrm{v}) - E(\psi_0).
\end{equation}
Figure \ref{JV_Energy_analysis} illustrates the nucleation energy per particle $E_{JV}/N$  as the function of the angular coordinate of the JV core.
In this scenario, the characteristics of the vortex energy are strongly influenced by the applied acceleration. 
The energy minimum occurs in the direction of the applied acceleration,  with the depth of the potential well, being proportional to the magnitude of the acceleration, $|\mathbf{a}|$.

The characteristics of the acceleration-induced effective potential well are reflected in the dynamics of the BEC, particularly in the azimuthal motion of the vortex along the barrier between rings, $\varphi_{JV}(t)$, as illustrated in Fig. \ref{protocol_and_potential}. The vortex and antivortex circulate in opposite directions at zero acceleration. However, as shown in the next section and illustrated in Fig. \ref{protocol_and_potential} (b), for the system with dissipation, the angular positions of both the vortex (solid red line) and the antivortex (dashed red line) ultimately align with the direction of the applied acceleration. By examining the features of this motion, one can extract both the direction and magnitude of the applied acceleration, thus offering a mechanism for acceleration sensing based on JV dynamics. 

\section{Relaxation dynamics of Josephson vortices driven by linear acceleration}

Dissipative effects are crucial in modeling nonequilibrium behavior, such as vortex nucleation, as they drive relaxation to equilibrium. 
Dissipation causes the vortex line to drift to the condensate edge (where vortices decay) or to pin it in the central hole of a ring-shaped condensate. 
Relaxation of the vortex core to the local energy minimum forms a metastable persistent current. 
In a trapped condensate, these effects arise from interaction with a thermal cloud and are phenomenologically described by the dissipative Gross-Pitaevskii equation (DGPE) \cite{Choi98,Pitaevskii1958}. 
For weakly interacting degenerate atoms near the thermodynamic equilibrium and under weak dissipation, the DGPE for the macroscopic wave function is given by
\begin{equation}
(i-\gamma)\frac{\partial\Psi}{\partial t}= \left[\hat{\mathcal{H}}-\mu\right]\Psi,
\label{eq:Dissip_GPE}
\end{equation}
where $\gamma\ll 1$ is the dissipation rate. 
Our main results are qualitatively independent of the chosen value of \(\gamma \). 
The dissipation rate \(\gamma\) determines the relaxation time of the system to a (meta)stable state: the larger \(\gamma\), the shorter the relaxation time.  \textcolor{black}{
In our dynamical simulations, the chemical potential $\mu(t)$ of the equilibrium state was adjusted at each time step to conserve number of particles, $N$.}  It is important to note that introducing phenomenological dissipation in this manner models the dissipation within an accelerating frame, where the thermal cloud is assumed to co-move with the condensate. A detailed discussion of physically relevant approaches for introducing phenomenological dissipation in the condensate under the influence of acceleration is provided in Ref. \cite{Dimer_accelerated}.


One of the key features of quantum vortices, as topologically protected states, is their inherent stability. Both, theoretical \cite{kanai2019merging,OLIINYK2020105113} and experimental \cite{pezze2024stabilizing, hernandez2024connecting} studies have demonstrated that a merging double-ring system exhibits the formation of spiral interference patterns associated with vortex flows. These patterns provide a robust and accessible method for measuring the angular momentum state of toroidal condensates. 

\begin{figure}[h!]
    \centering
\includegraphics[width=1.0\linewidth]{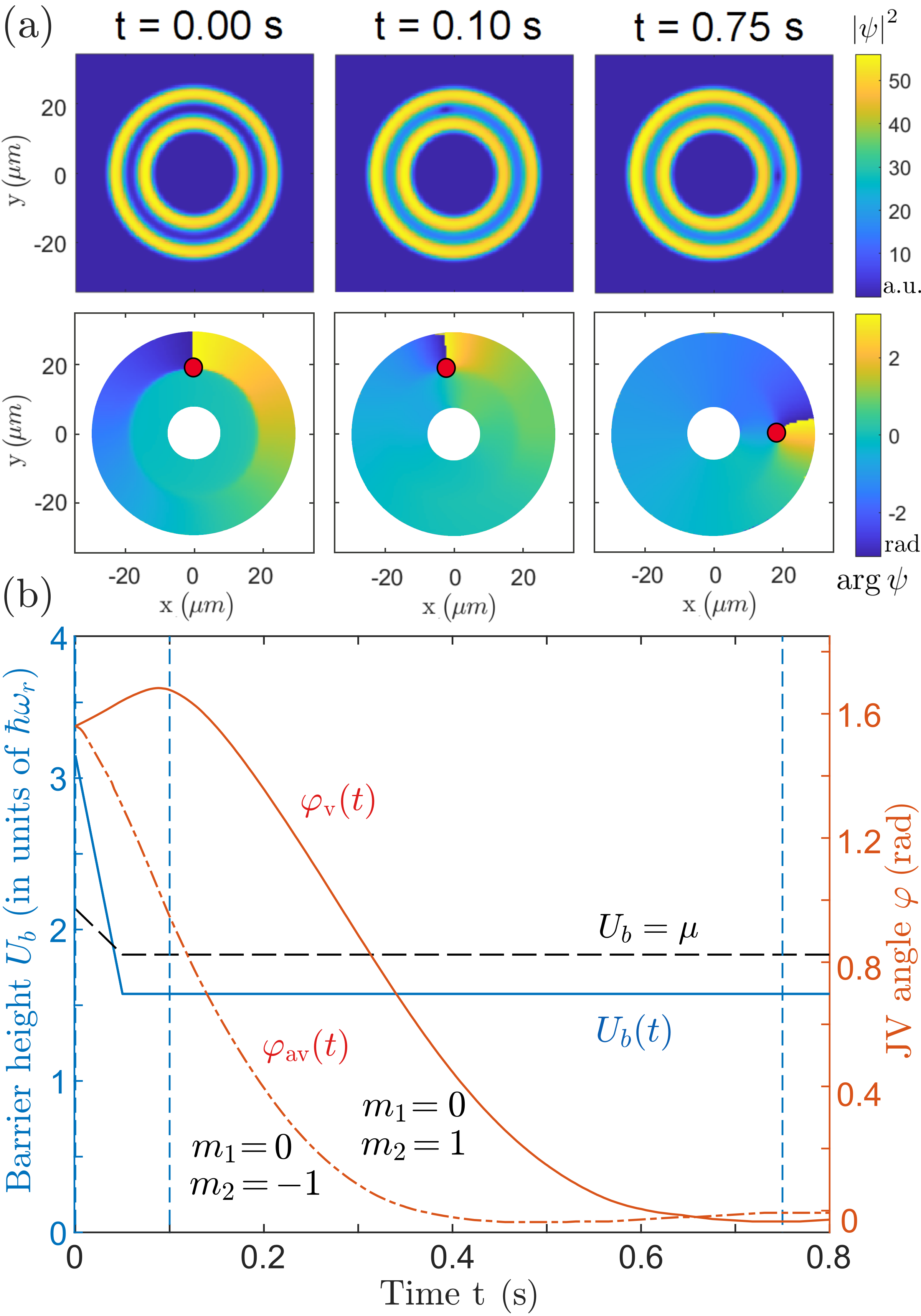}
\caption{(a) Dynamics of a single JV (red circle indicates position of the vortex core) under constant horizontal acceleration $a = \SI{2}{\milli\meter/\second^2}$ ($\varphi_a = 0$),  $l_b = \SI{1.43}{\micro\meter}$, and $N = 5 \times 10^4$, showing density (upper row) and phase (\textcolor{black}{lower} row) evolution. The JV aligns with the acceleration, stabilizing at $\varphi = \varphi_a = 0$ due to nonzero dissipation $\gamma=0.015$. (b) Barrier amplitude $U_b(t)$ and JV angular coordinate, $\varphi(t)$, vs time. Starting at $t=0$, the barrier amplitude $U_b$ is ramped down over 0.05~s from $1.5\,\mu_0$ to $0.75\,\mu_0$ \textcolor{black}{, where $\mu_0 = \mu(0)$,} and then remains unchanged (blue solid line). The red lines show the JV's angular coordinate over time for vortex (red solid line) and anti-vortex (red \textcolor{black}{dash-dotted} line). The black dashed line denotes the barrier amplitude at the chemical potential level, \textcolor{black}{$U_b = \mu(t)$}. }
\label{protocol_and_potential}
\end{figure}

In this work, we investigate the behavior of a double-ring system subjected to constant linear acceleration, which breaks the symmetry of the system and consequently leads to an asymmetric arrangements of the tunneling currents. It is therefore expected, that locations of vortex cores inside the circular barrier can be explicitly related to the direction and magnitude of the acceleration. In our simulations we use unbiased ($\Delta\mu=0$) double ring-system with barrier \textcolor{black}{amplitude} $U_b>\mu$, as the initial condition. To detect the vortex positions in both, the density and phase distributions, we reduce the barrier amplitude to a constant value just below the chemical potential, as shown in Fig. \ref{protocol_and_potential}, with a time constant long enough to suppress the formation of spiral structures. As a result of the relaxation process, we observe stationary vortex positions that exhibit sensitivity to the applied acceleration.

\begin{figure}[h!]
    \centering
    \includegraphics[width=1.0\linewidth]{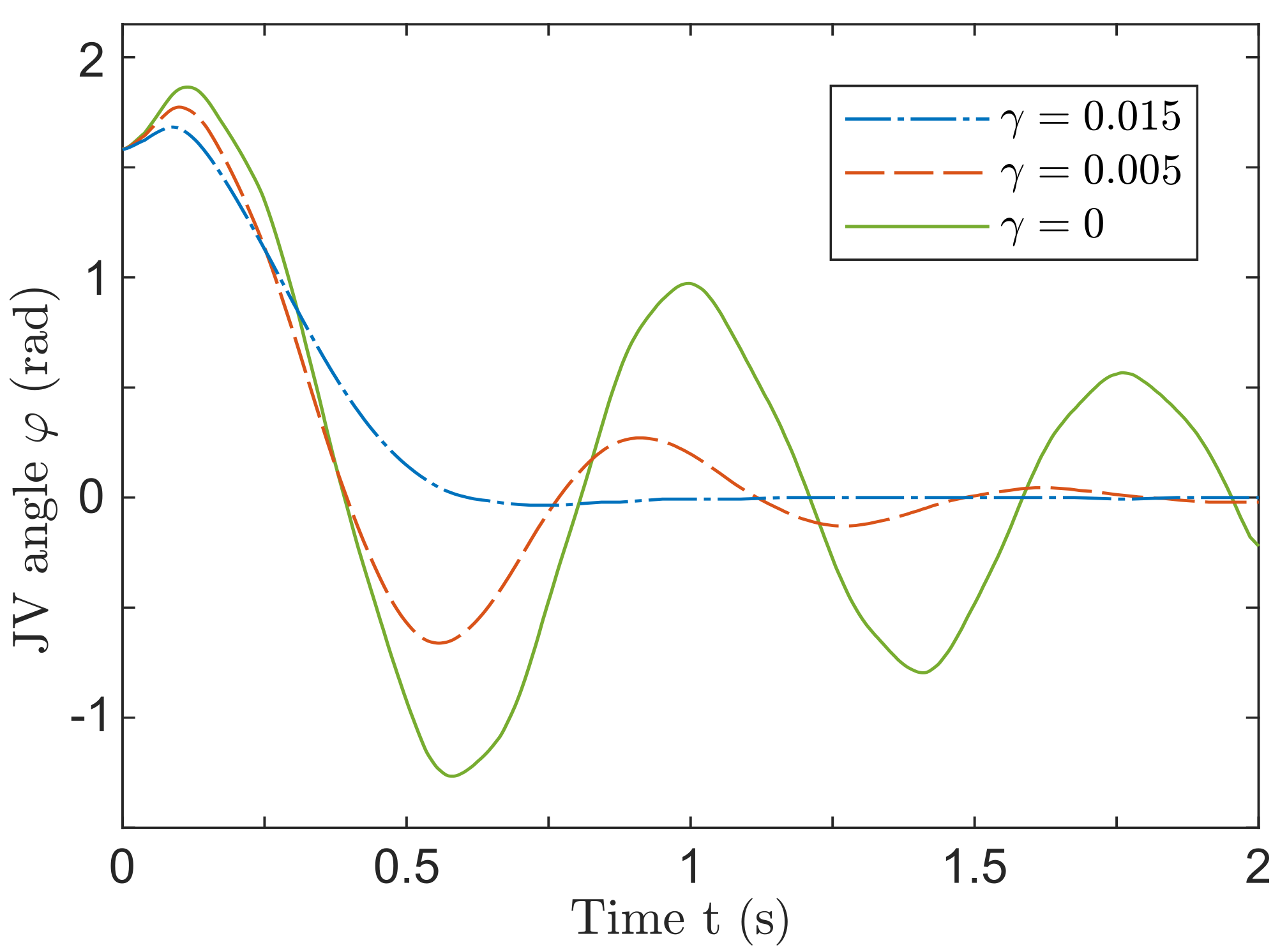}
    \caption{Angular dynamics of a single JV under constant acceleration $a = \SI{2}{\milli\meter/\second^2}$ for different dissipation rates $\gamma$, with $\varphi_a = 0$. Higher $\gamma$ reduces the \textcolor{black}{relaxation} time of the JV at $\varphi = \varphi_a$, aligned with the acceleration for $l_b = \SI{1.43}{\micro\meter}$, $N = 5 \times 10^4$.
    }
    \label{fig:Single_JV_dissipation_role}
\end{figure}

Linear acceleration induces an azimuthal asymmetry in the density distribution, making the angular position aligned with the acceleration direction energetically favorable for the JV, as illustrated in Fig.~\ref{JV_Energy_analysis}. Notably, the local energy minima deepen with increasing acceleration. Consequently, in a conservative system, the JV either circulates within the lower-density region between the rings at low acceleration rates or undergoes periodic oscillations along the direction of acceleration when the acceleration is sufficiently high for a given chemical potential difference.
In realistic experiments at finite temperature, interactions between the condensate and the thermal cloud introduce dissipative effects, which cause the JV's angular position to eventually align with the direction of acceleration after a relaxation period. An additional mechanism contributing to the decay of JV oscillations, even for $\gamma = 0$, is the emission of acoustic waves during vortex drift in an inhomogeneous condensate \cite{Parker2004}. Figure \ref{fig:Single_JV_dissipation_role}  illustrates the evolution of the JV's angular position in both conservative and dissipative regimes. As expected, increasing the dissipation rate $\gamma$ accelerates the relaxation process.

\begin{figure}[h!]
    \centering
    \includegraphics[width=1.0\linewidth]{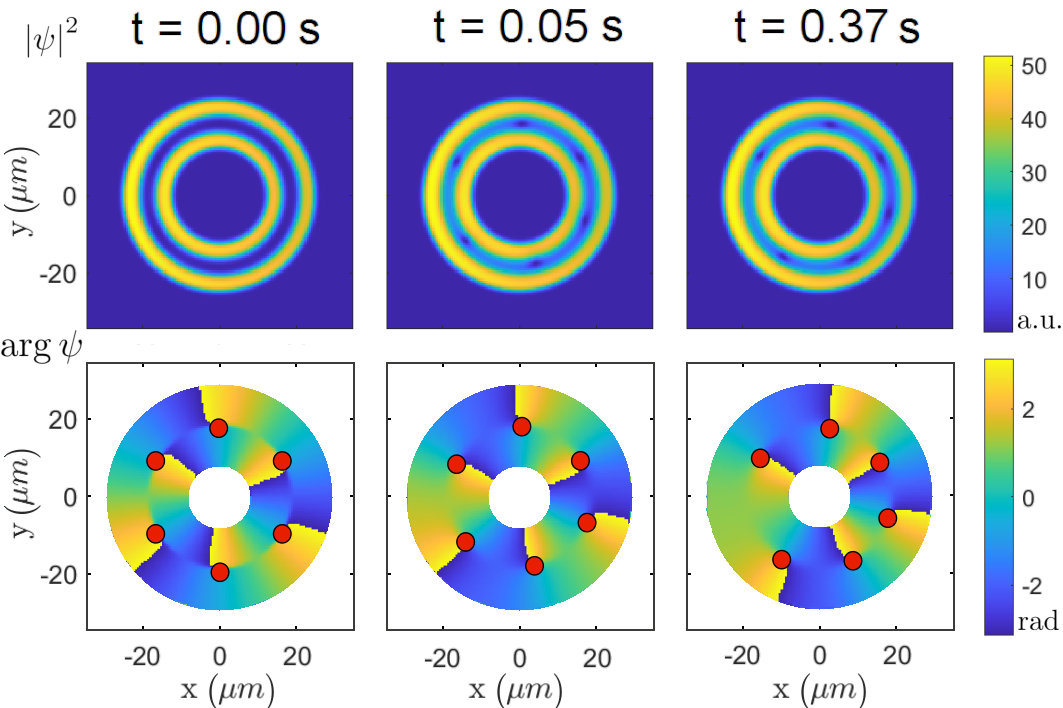}
    \caption{Density (upper row) and phase (lower row) snapshots of two coaxial rings with six JVs under a constant horizontal acceleration of $a=\SI{2}{\milli\meter/\second^2}$ and dissipation rate $\gamma=0.015$ at various times after lowering the barrier height. JV cores are marked by red circles. Initially symmetric, the JV lattice is deformed by acceleration, stabilizing into an asymmetrical configuration. During the first $t=0.05, \text{s}$, the inter-ring barrier is linearly lowered to facilitate imaging of JVs and pinning them in the radial direction.}
    \label{fig:dynamics_JV6}
\end{figure}

We quantify the asymmetry of the JV lattice by the relative deviation of its centroid \textcolor{black}{$d = |\mathbf{d}|$}, which describes the collective displacement of all vortices:
\begin{equation}
    \mathbf{d} = \frac{1}{{R_b}N_{JV}}\sum_{n=1}^{N_{JV}}{{\mathbf{r}_n}},
\end{equation}
where $\mathbf{r}_n$ is the radius vector of the $n$-th vortex, $N_{JV}$ is the total number of JVs, and $R_b$ is the radius of the circular barrier, where the JVs are located. 

In the absence of acceleration, JVs are symmetrically positioned at the vertices of a regular $n$-gon, resulting in zero relative deviation ($d = 0$). 
When acceleration is applied, all JVs shift in the direction of the acceleration, $\mathbf{d} \!\parallel\! \mathbf{a}, ~ \textcolor{black}{d=|\mathbf{d}| > 0}$, finding a new equilibrium position influenced by mutual repulsion (as all vortices have the same sign). 
After equilibration, the system exhibits an intermediate asymmetry $0 \le d < 1$, pointing towards the direction of acceleration. A typical example of the evolution of a lattice formed by six JVs is shown in Fig.~\ref{fig:dynamics_JV6}.



\begin{figure}[h!]
    \centering
    \includegraphics[width=1.0\linewidth]{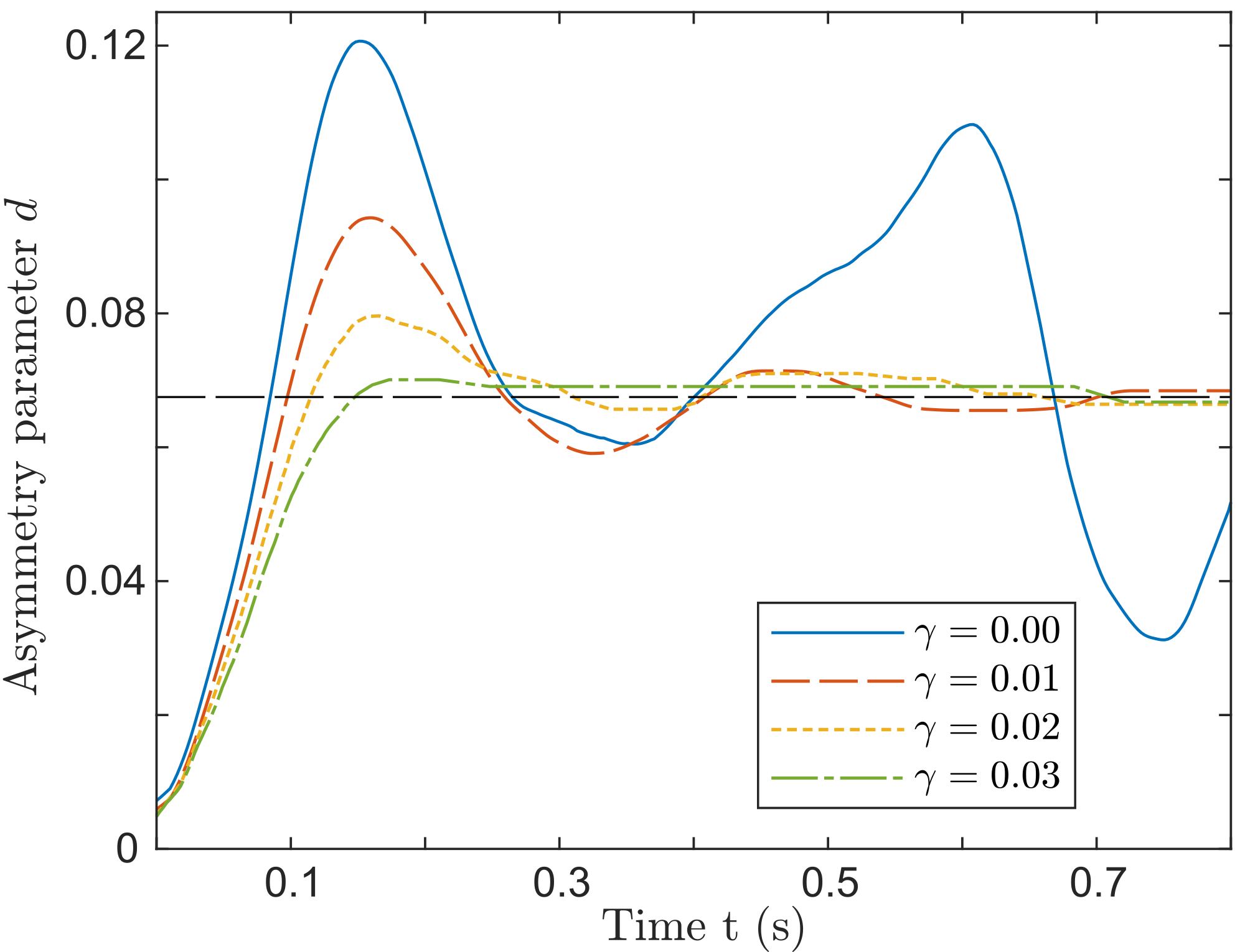}
    \caption{Evolution of the asymmetry parameter $d$ for six JVs under constant acceleration $a = \SI{2}{\milli\meter/\second^2}$, 
     $l_b = \SI{1.43}{\micro\meter}$, $N = 5 \times 10^4$ for various dissipation rates $\gamma$. While $\gamma$ affects the equilibration dynamics of $d$ - exhibiting decaying oscillations for $\gamma \leq 0.025$ and becoming aperiodic for $\gamma > 0.025$ - the equilibrium asymmetry value remains independent of the dissipation rate.}
\label{Deformation_is_independent_from_dissipation}
\end{figure}

The equilibration of the system with a non-zero dissipation rate $\gamma$ enables us to determine the actual value of the asymmetry parameter $d$ in the equilibrium state of the JV chain. 
It is essential to ensure that the equilibrium value of $d$ is independent of the specific dissipation rate used. 
To verify this, we compared the evolution of six JVs with different dissipation rates from $\gamma=0$ to $\gamma=3\cross 10^{-2}$ and confirmed that the final equilibrium value of $d$ does not depend on $\gamma$ (see Fig.~\ref{Deformation_is_independent_from_dissipation}). 
For our simulations, we use an intermediate value of $\gamma = 1.5\cross 10^{-2}$ for dissipation.

\begin{figure}[h!]
    \centering
\includegraphics[width=1.0\linewidth]{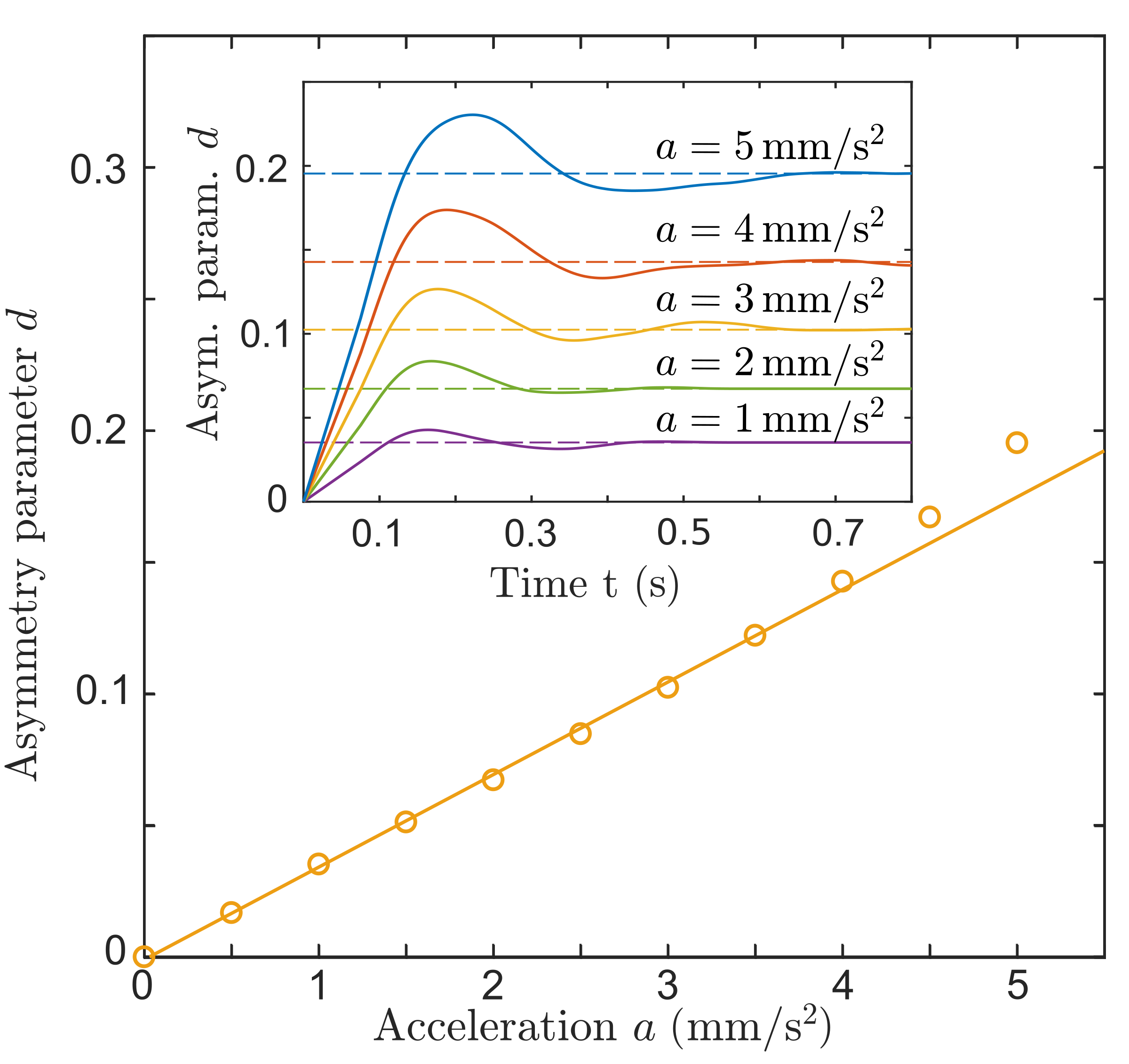}
    \caption{Equilibrium asymmetry parameter $d$ as a function of linear acceleration $a$, illustrating a linear dependence at low accelerations for $l_b = \SI{1.43}{\micro\meter}$, $N = 5 \times 10^4$. (Inset) Time evolution of the asymmetry parameters for varying accelerations $a$ (solid lines) with their respective equilibrium values for $\gamma = 0.015$ (dashed lines).
    }
    \label{Stable_deformation}
\end{figure}

The equilibrium asymmetry parameter $d$ shows a linear dependence on the applied acceleration $a$, following the relation $d = 3.52 \times 10^{-2} \, \text{s}^2/\text{mm} \times a$ (see Fig.~\ref{Stable_deformation}) for small accelerations. However, for \textcolor{black}{$a \gtrsim \SI{4.5}{\milli\meter/\second^2}$, a deviation from this linear behavior is observed, likely due to the significant asymmetric density bias in the barrier region.
As typically intended of quantum sensing devices, low acceleration rates are accessible to measurements, with the sensitivity of this method being limited at very low accelerations as the variation in the asymmetry parameter approaches the detection resolution limits of the JV core positions.} Conversely, measuring higher acceleration rates necessitates longer relaxation times, which prolongs the measurement process. Additionally, at very high acceleration rates, density biases may cause the ring-shaped condensate to break apart, further restricting the maximum accessible acceleration rate.


\section{Conclusions}
In the present work we have considered the AC Josephson effect in coaxial two-dimensional ring-shaped condensates separated by a potential barrier and investigated the dynamics of Josephson vortices within such a double-ring BEC. Through direct simulations of the Gross-Pitaevskii equation, we have analyzed tunnelling superflows driven by an initial imbalance in atomic populations of the rings. The superflows through the Bose-Josephson junction are strongly influenced by persistent currents in the concentric rings, leading to pronounced Josephson oscillations in population imbalances for co-rotating and non-rotating states. The azimuthal configuration of the tunneling flow requires  the formation of Josephson vortices, resulting in zero net current through the junction for rings with different angular momentum states. \textcolor{black}{However, if a linear acceleration is applied to the system and there is only one vortex in the junction, the population imbalance oscillations can be restored.}


A key aspect of this study is exploring how linear acceleration affects the dynamics of Josephson vortices. We have found that acceleration leads to an asymmetric displacement of vortices, an effect that can be utilized to determine both the magnitude and direction of the acceleration. By \textcolor{black}{introducing an} asymmetry parameter for the vortex lattice after equilibration, we have demonstrated that this parameter is linearly proportional to the absolute value of the applied acceleration. We have proven that the equilibrium angular position of a single vortex is sensitive to the direction of acceleration. 


\textcolor{black}{These findings provide new insights into the Josephson effect in a bosonic junction modified by linear acceleration and introduce prospective methods  for quantifying acceleration effects via analysis of interference patterns experimentally observed in atomtronic systems \cite{PhysRevA.108.063306,hernandez2024connecting,PhysRevX.12.041037}.}

\begin{acknowledgments}
The authors thank Luca Salasnich,  Yelyzaveta Nikolaieva, Mark Edwards, and Yuriy Bidasyuk for useful discussions.  YB, OP, NB, and AY
acknowledge support from the National Research Foundation of Ukraine through Grant No. 2020.02/0032. AY acknowledge support from the project ‘Theoretical analysis of quantum atomic mixtures’ of the
University of Padova and from INFN.
\end{acknowledgments}

\bibliography{ref}

\end{document}